\documentclass{aanew}
\usepackage{psfig}

\def\bron{SAX~J2239.3+6116}
\def\ecs{erg~cm$^{-2}$~s$^{-1}$}

\def\lum{erg~s$^{-1}$}

\begin{document}

\title{Discovery of a 1247~s pulsar in the Be X-ray binary \bron}
\titlerunning{Discovery of a 1247~s pulsar in \bron}
\author{J.J.M.~in~'t~Zand\inst{1,2}
\and J.~Swank\inst{3}
\and R.H.D.~Corbet\inst{3,4}
\and C.B.~Markwardt\inst{3,5}
}
\offprints{J.J.M.~in~'t Zand (at e-mail {\tt jeanz@sron.nl})}

\institute{     Astronomical Institute, Utrecht University, P.O. Box 80000,
                NL - 3508 TA Utrecht, the Netherlands
         \and
                SRON National Institute for Space Research, Sorbonnelaan 2,
                NL - 3584 CA Utrecht, the Netherlands
     \and
                NASA Goddard Space Flight Center, Code 662, Greenbelt,
                MD 20771, U.S.A.
     \and
                Universities Space Research Association, U.S.A.
     \and
                University of Maryland, Department of Astronomy, College Park, MD, 20742, U.S.A.
        }
\date{Received, accepted }

\abstract{
A search for pulsations from the Be X-ray binary \bron, through observations
with the Narrow Field Instruments on BeppoSAX and the Proportional Counter
Array on RXTE, yielded the clear presence of a 1247~s coherent oscillation.
Given the fairly high X-ray luminosity on previous occasions (up to a few
times 10$^{36}$~erg~s$^{-1}$ in 2 to 28 keV), the oscillation must be due to
the spin of a neutron star. Assuming that the 262 day recurrence time is the
orbital period, \bron\ has both the longest orbital period and the longest
pulse period of 24 Be X-ray binaries with measured orbital periods. 
}

\maketitle

\section{Introduction}
\label{intro}

\bron\ was discovered as a moderately bright X-ray transient through
serendipitous observations with the Wide Field Cameras (WFCs) on BeppoSAX in
1997 and 1999 (In~'t~Zand et al. 2000). A subsequent archival search of 
data from the All-Sky Monitor (ASM) on RXTE showed four unambiguous
outbursts from the same source, one of which is simultaneous with a
WFC detection. The five outbursts thus observed occurred with regular intervals
of $262\pm5$~d. The highest X-ray flux ever measured was sampled with the WFCs
at 1~10$^{-9}$~\ecs\ (2-28 keV). A 20~ksec exposure with the Narrow
Field Instruments on BeppoSAX in 1998 resulted in 
a more accurate X-ray position.
Optical follow-up observations of this error box
revealed a candidate optical counterpart: a Be star at an
estimated distance of 4.4~kpc. The combination of periodicity, peak flux
and optical counterpart prompted In~'t~Zand et al. (2000) to classify \bron\
as a Be X-ray binary (BeXB) with an orbital period of 262~d. Pulsations,
a common attribute of a BeXB, were not found but the limits were not very
constraining, simply because no sensitive X-ray measurements had been
obtained during an outburst.

High-mass X-ray binaries (combining the catalog of Liu, van Paradijs \& van
den Heuvel 2000 with more recent results) include 24 BeXBs with measured
orbital periods $P_{\rm o}$.
Interestingly, \bron\ is the case with the longest $P_{\rm o}$, followed by
X Per (250.3~d; Delgado-Mart\'{\i} et al. 2001) and GS~1843-024
(242.18~d; Finger et al. 1999).
It is expected, based on the $P_{\rm o}$/$P_{\rm p}$ correlation found in the
'Corbet' diagram (Fig.~\ref{figcorbet}), that if \bron\ contains a pulsar it
should have a pulse period near $10^3$~s. A non-detection would bring a black
hole hypothesis into serious consideration (like for 4U~1700-377, see Brown
et al. 1996). 

\begin{figure}[t]
\psfig{figure=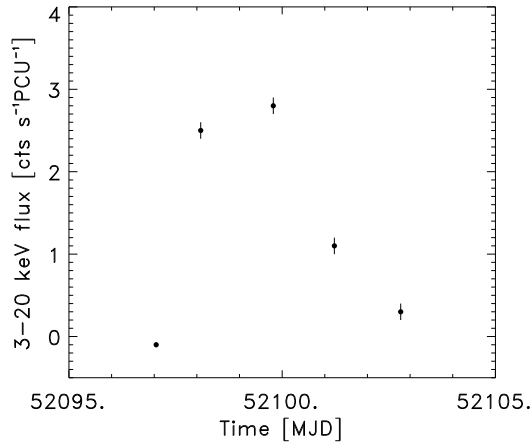,width=0.9\columnwidth,clip=t}

\caption[]{Light curve generated from the 5th through 9th PCA measurement.
1~cts~s$^{-1}$PCU$^{-1}$ is equivalent to 0.5 mCrab or
0.9~10$^{-11}$~\ecs.
\label{figpca}}
\end{figure}

\begin{figure}[t]
\psfig{figure=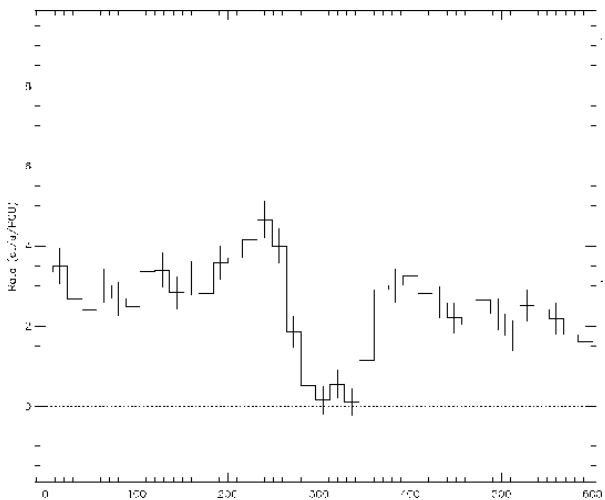,width=0.9\columnwidth,clip=t,angle=270.}

\caption[]{Background-subtracted light curve from the 6th PCA observation.
The dotted line refers to the zero-flux level.
\label{figpca2}}
\end{figure}

We set out to search for the X-ray pulsar with sensitive measurements,
at times when the transient is active.  The first opportunity to do so was
presented
by the fifth BeppoSAX Announcement of Opportunity in which we applied for
target-of-opportunity observations at the earliest occurring outburst.

\section{Observations, light curve and spectrum}
\label{secobs}

The ephemeris as determined from the ASM data predicted an outburst on July
7, 2001 (MJD~52098), with an uncertainty of approximately 10 days. Since the
anticipated peak flux was too low for ASM observations to be able to trigger
the BeppoSAX observation, public snapshot observations were initiated with
the sensitive Proportional Counter Array (PCA) on RXTE on June 30, 2001.
The PCA (e.g., Jahoda et al. 1996) consists of five non-imaging Proportional
Counter Units (PCUs) with collimators that limit the field of view to 1\degr\
full-width at half maximum (FWHM). The collecting area is 6500~cm$^2$ and the
bandpass 2 to 60 keV. The snapshot observations were performed roughly every
two days. Nine measurements were carried out in total with exposure times
between 0.78 and 2.10 ksec, some with 5 PCUs operating, and some with only
2 PCUs. The last one was on July 12. Fig.~\ref{figpca} shows the light
curve generated from part of these data. \bron\ was first detected in the sixth
snapshot observation on July 8th. In Fig.~\ref{figpca2} we show the light curve
of this 6th observation with 16~s time resolution. The light curve shows
considerable variability without a clear periodicity. Most notably, there is
a 80~s long dip during which the flux drops to near zero. The mid-time of
the dip is MJD~52098.09305$\pm0.00012$. For the 7th and 8th observation there
was also similar variability although less definitive because fewer
detectors were on.

The PCA detection triggered our TOO with the BeppoSAX Narrow Field Instruments
(NFI). The NFI consist of
4 devices. The Low-Energy and Medium-Energy Concentrator Spectrometers
(LECS and MECS) are imaging devices in the 0.1 to 10 keV band
(Parmar et al. 1997 and Boella et al. 1997). \bron\ was not unambiguously
detected in the other two devices, consistent with their sensitivities.
The observation was carried out between July 12.803 and 13.851 UT
(MJD 52102.803 through 52103.851).
The net exposure times are 40,039~s for the MECS and 17,952 for the LECS.
We analyzed the data in a standard manner, by accumulating photons
within 4\arcmin\ (MECS) or 8\arcmin\ (LECS) from the source centroid.
We applied a background correction by employing
long-duration data from high latitude fields, grouped energy channels
such that at least 20 photons are present in each bin, and used
standard response matrices. The validity of the background
model was verified through a comparison of both data sets (i.e.,
the high latitude and the source data sets) for photons further than
5\arcmin\ and 10\arcmin\ from the source centroid for the MECS and LECS
respectively. The MECS detected $1610\pm43$ photons
from \bron\ and the LECS $149\pm24$. For the spectral analysis, the bandpasses
were restricted to the photon-rich and well-calibrated ranges of 1.0-4.0 keV
for the LECS and 1.6-10.5 keV for the MECS. The data could be fitted
satisfactorily with a
simple absorbed power law model ($\chi^2_{\nu}=1.19$ for $\nu=49$ degrees
of freedom). The photon index is $1.6\pm0.2$, the absorbing column density
$N_{\rm H}=(3\pm1)~10^{22}$~cm$^{-2}$, and the average 2 to 10 keV flux
$3.8~10^{-12}$~\ecs\ ($4.7~10^{-12}$~\ecs\ unabsorbed). Over the course of
the observation the source showed a decreasing trend with a flux reduction
of $20\pm5$\%.

Observation 6 of the PCA gives a spectrum for which the 2.5-30 keV
range can be fit acceptably with a power law or a power law with a
high energy cut-off. If the column density is fixed at the value seen
with the LECS and MECS, the power law photon index when the source was
seven times brighter was 1.1$\pm$0.2, with a high energy cut-off
beginning at 6.3$\pm$0.9 keV and e-folding energy of 7$\pm$1 keV.
The 2-10 keV flux is $2.3~10^{-11}$~\ecs. The spectral shape is typical
for an accretion-powered pulsar (e.g., White et al. 1983, Mihara 1995,
Dal Fiume et al. 2000).

\begin{figure}[t]
\psfig{figure=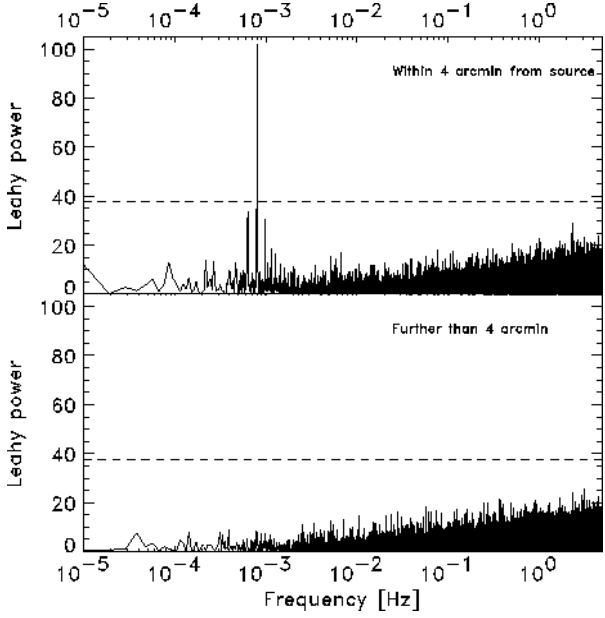,width=0.9\columnwidth,clip=t}

\caption[]{Fourier power density spectrum from all 1855 MECS-detected
photons within 4\arcmin\ from the source (top) and from the remaining
9848 photons (bottom). The power is defined following Leahy et al. (1983).
The dashed lines represent the 99\%-confidence detection levels. 
\label{figfps}}
\end{figure}

\section{Detection of oscillations}

\begin{figure}[t]
\psfig{figure=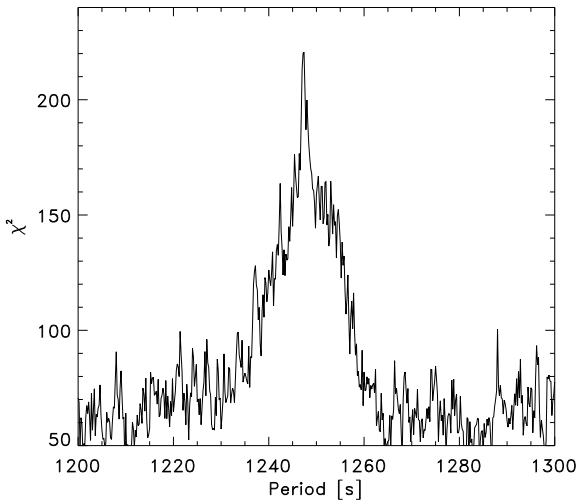,width=0.9\columnwidth,clip=t}

\caption[]{Periodogram of source photons (same data as used for
spectrum presented in the top panel of Fig.~\ref{figfps}). A
phase resolution of 64 was employed.
\label{figperiodogram}}
\end{figure}

\begin{figure}[t]
\psfig{figure=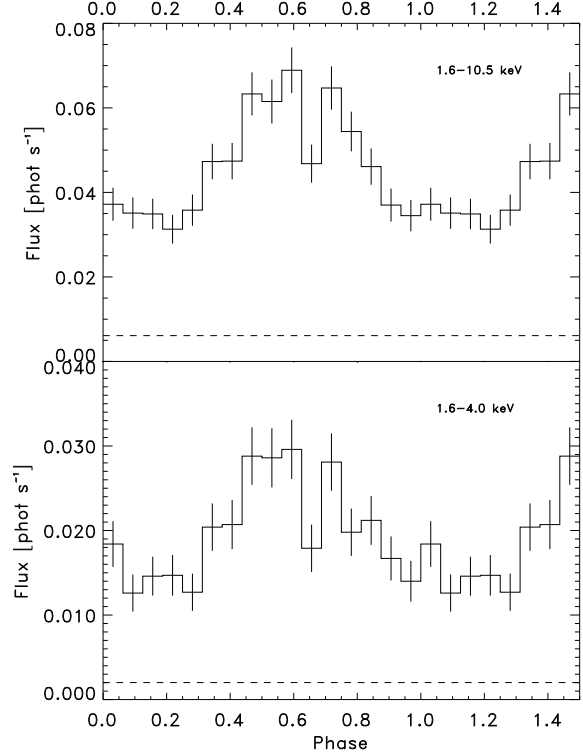,width=0.9\columnwidth,clip=t}

\caption[]{Light curve in two bands, folded at the period of 1247.2~s.
Phase is defined to be zero for the start of the observation. The dashed
lines show the background levels.
\label{figprofile}}
\end{figure}

Fig.~\ref{figfps} shows the Fourier power density spectrum of all MECS
photons (i.e., without background subtraction), both from within the source
point-spread function and outside. A clear peak
is present at 0.80~mHz, with sidelobes separated by 0.18~mHz from the
main peak due to aliasing through occultations by the Earth during
each BeppoSAX orbit and passages through the South-Atlantic Geomagnetic
Anomaly. As a check on the reality of the oscillation, we
confirmed that the  non-source photons of the same observation show
no power at the relevant frequency.
A periodogram of the same data for the source photons, see
Fig.~\ref{figperiodogram}, gives the most accurate measurement of the
period: $1247.2\pm0.7$~s. We identify this as a pulsar signal.

Fig.~\ref{figprofile} shows the light curve folded with this period. There
is a broad peak and the modulation depth is $60\pm10$\%. A conspicuous
feature is a dip at phase 0.65. The epoch of this dip is
MJD~52102.81318$\pm0.00045$. The PCA data show a dip of similar
duration but larger depth (Fig.~\ref{figpca2}). The time difference
between both dips is $326.988\pm0.036$ pulse cycles which is consistent with
both dips having identical pulse phases. PCA observations 7 and 8 show similar
dips at similar phases but with different depths. Assuming the dips are
phase locked, we can refine the pulse period to $1247.15\pm0.14$~s.
Features likes this dip are not uncommon in pulse profiles of accretion-powered
X-ray pulsars, particularly at energies below 10 keV (for recent overviews of
pulse profiles, see Mihara 1995 or Bildsten et al. 1997).

\section{Discussion}

\begin{figure}[t]
\psfig{figure=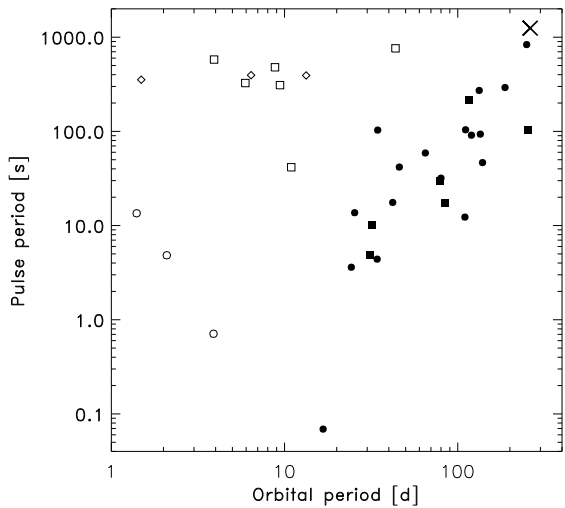,width=0.9\columnwidth,clip=t}

\caption[]{Corbet diagram, updated with SAX J2239.3+6116 (cross).
Open symbols are supergiant systems, filled symbols
Be systems. For the Be systems, the circles indicate optically identified
cases and squares those that are not. The open squares refer to optically
identified Roche-lobe underfilling supergiants and the open circles to
Roche-lobe filling giants. The diamonds refer to systems without optical
counterpart.
\label{figcorbet}}
\end{figure}

The ASM data between January 1996 and July 2001 encompass eight predicted
times of outbursts, but only the first four are readily detectable
above a threshold of about 1 mCrab. The 262~d orbital period was basically
determined from those four ASM detections (In~'t~Zand et al. 2000). The
fifth outburst detected with WFC is consistent with that periodicity.
Given the lack of clear outbursts afterwards, one might perhaps wonder whether
the 262~d periodicity is well defined and can be identified with the orbital
period. Quasi-periods of similar magnitude and variable peak flux have been
found in other X-ray binaries without being associated to the orbit (e.g., in
Terzan 6, see In 't Zand et al. 1999). We argue that the periodicity in
\bron\ is indeed the orbital period, because the coherence of the periodicity
is rather high and because orbital periods of this magnitude are expected for
BeXBs and the optical Be-star identification is rather secure (the estimated
chance probability is less than 10$^{-4}$ which is low for such a high-flux
source, see also In 't Zand et al. 2000) . The fact that the outbursts have
become harder to see in the ASM data is very likely due to them becoming
significantly shorter: the average duration of the first four outbursts was
about 1 month, while the July 2001 outburst appears to have lasted just a
few days. The orbital period using the timing
of the July 2001 outburst reveals a refined value of $P_{\rm o}=262.6\pm0.7$
which is close to what was measured previously but more precise.

Could the compact object be a white dwarf rather than a neutron star? The
distance of 4.4~kpc (In~'t~Zand et al. 2000) implies that the maximum 2-28 keV
luminosity ever measured was 2.3~10$^{36}$~\lum. This is 3 orders of magnitude
above what is normal for non-synchronously rotating magnetic white dwarfs
accreting from a companion star (e.g., Eracleous et al. 1991 and Patterson
1994). A number of white dwarf systems have
been observed with $\sim10^{36}$~\lum\ luminosities or higher, but those have
supersoft spectra (with $kT$ of order a few tens of eV; e.g., G\"{a}nsicke
et al. 2000). \bron\ did not, and we dismiss the white dwarf scenario.

The variably-peaked (1.5~mCrab
for the July 2001 outburst and 20~mCrab for the May 1997 outburst\footnote{one
should keep in mind that peak fluxes may be biased 
because of sparse sampling}) and periodic outburst behavior are typical
for Type I BeXB outbursts (e.g., Stella et al. 1986 and Bildsten et al. 1997)
which are explained as due to increased mass accretion rates by the
compact object when, during periastron in an eccentric orbit, it passes
through denser parts of the companion wind without forming
an accretion disk.

The values of spin period and orbital period for accretion-powered
pulsars with high-mass companion stars have been shown to qualitatively
reflect the nature of the stellar wind and the history of the neutron-star's
interaction with it (Corbet 1984; Waters \& van Kerkwijk 1989).
Figure~\ref{figcorbet} shows the position of \bron\ in the
$P_{\rm p}$--$P_{\rm o}$ diagram
for the periods 1247 s and 262 d. It would be at the long period end
of the distribution of pulsars with Be star companions, where that
distribution meets the distribution of supergiants. Since this system
has faint X-ray outbursts for only a part of a long orbital period,
determination of orbital parameters through Doppler shifts of the spin
period will be difficult. Pulse period changes are easier
to measure. Given the accuracy obtained for the July 2001 outburst,
spin-up time scales of less than 10$^3$~yr are measurable between two
outbursts. This would cover time scales seen in some systems
which can be as short as 10 to 100 years (Bildsten et al. 1997).

\begin{acknowledgements}
We thank the BeppoSAX Time Allocation Committee for the installation of
a provisional AO5 target list before final review, which
made possible the observation here discussed. Also, we thank the staff
at the BeppoSAX Science Operation Center and Data Center for guidance of
this observation. JZ acknowledges financial support from the Netherlands
Organization for Scientific Research (NWO). This research has made use of
SAXDAS linearized and cleaned event files (Rev.2.1.4) produced at the
BeppoSAX Science Data Center. {\em BeppoSAX\/} is a joint Italian and
Dutch program.

\end{acknowledgements}

\end{document}